\documentclass[journal,comsoc]{IEEEtran}
\IEEEoverridecommandlockouts
\normalsize

\ifCLASSINFOpdf

\else

\fi
\usepackage{soul}
\usepackage{array}
\usepackage{color}
\usepackage{amsfonts}
\usepackage{amssymb}
\usepackage{amsmath}
\usepackage[pdftex]{graphicx}
\usepackage{cite}
\usepackage{balance}
\usepackage{caption}
\usepackage{subcaption}
\usepackage{textcomp}
\usepackage{gensymb}
\usepackage{float}
\usepackage{comment}
\usepackage{subcaption}
\usepackage{tabularx}
\usepackage{multirow}
\usepackage{amsthm}
\usepackage{algorithmic}
\usepackage{algorithm}

\newcommand{\yvec}{\mbox{\boldmath $y$}}
\newcommand{\hvec}{\mbox{\boldmath $h$}}

\newcommand{\xvec}{\mbox{\boldmath $x$}}
\newcommand{\wvec}{\mbox{\boldmath $w$}}

\usepackage[utf8]{inputenc}
\begin{document}

\title{Digital Twin Enhanced Channel Twin for AI-Native CSI Inference: Generalizability and Scalability}

\author{Majumder Haider, ~\IEEEmembership{Graduate Student Member,~IEEE}, Imtiaz Ahmed, ~\IEEEmembership{Senior Member,~IEEE}, Zoheb Hassan, ~\IEEEmembership{Member,~IEEE}, Danda B. Rawat, ~\IEEEmembership{Senior Member,~IEEE}, and Huaiyu Dai, ~\IEEEmembership{Fellow,~IEEE}}

\maketitle

\begin{abstract}
Accurate channel state information (CSI) is critical for advanced multi-antenna wireless networks. While  high-fidelity and site-specific ray-tracing (RT) equipped wireless digital twins can overcome overhead for CSI acquisition. However, computing deterministic, calibrated RT based CSI from a wireless digital twin for every orthogonal frequency-division multiplexing (OFDM) symbol violates the strict microsecond latency budgets of the 5G NR numerology. To overcome this computational bottleneck, this paper investigates three approaches within a calibrated 3D digital twin framework, namely (i) the generalization of the channel twin, (ii) the performance enhancement using a neural receiver, and (iii) a data-driven interpolation framework for scalability. We compute high-precision CSI for a sparse subset of temporal anchors and employ an attention-based Transformer to predict the remaining symbols. Unlike polynomial splines or sequential long short-term memory (LSTM) networks, the Transformer exploits a global receptive field to capture the non-linear multipath dynamics while enabling parallelizable, real-time inference. To achieve spatial scalability, we introduce channel twin generalization. By fine-tuning the 3D RT models alongside the Transformer, the framework leverages the CSI dataset of one location to infer the channel behavior of an unseen environment. Transfer learning further adapts the model to a new environment using only a small fraction of locally collected data. Simulation results demonstrate that the proposed architecture substantially outperforms the baseline interpolators, achieves robust spatial transferability, and lowers the bit error rate through the unified neural receiver. These results establish a scalable, environment-agnostic foundation of distributed channel twins for high-quality, low-overhead CSI acquisition  in next-generation cellular networks.
\end{abstract}


\section{Introduction}
\subsection{Introductory Background}
The 6G vision, guided by the ITU IMT-2030 framework, envisions an intelligent and autonomous ecosystem that integrates communication, sensing, and computing \cite{IMT2030}. Beyond raw data transmission, 6G is expected to natively embed artificial intelligence (AI) into the network, so that the infrastructure operates simultaneously as a communication link and a high-resolution sensing system. The associated performance targets are stringent. The framework targets peak data rates on the order of 1 terabit per second (Tbps), user-experienced rates of 1 to 10 Gbps, over-the-air latency between 10 and 100 microseconds, and a connection density of up to ten million devices per square kilometer \cite{IMT2030}. These targets impose tight constraints on the physical layer and motivate a shift toward AI-native air interfaces.

Within this vision, accurate channel state information (CSI) becomes a foundational enabler rather than a routine link-optimization task. High-fidelity CSI is essential for the aggressive beamforming and spectral efficiency (SE) required by massive multiple-input multiple-output (MIMO) systems. It is equally critical for integrated sensing and communication (ISAC), an emerging feature of the early 6G frameworks of the 3rd Generation Partnership Project, where a single waveform must support high-rate data transmission and high-resolution environmental sensing. This requirement becomes even more acute at the sub-Terahertz (sub-THz) frequencies and ultra-massive MIMO arrays envisioned for 6G. At these frequencies, radio waves suffer from severe propagation loss and are highly sensitive to blockages. Maintaining a stable link therefore depends on narrow, highly directive beams, which in turn depend on real-time, high-fidelity channel estimates.

Acquiring such CSI in dynamic environments is, however, a critical bottleneck. Legacy methods of estimating and feeding back the CSI generate signaling overhead that does not scale to the array sizes and bandwidths of 6G. As a result, the standardization effort is pivoting toward AI-native air interfaces that predict, compress, and reconstruct CSI with low overhead. In parallel, site-specific wireless digital twins have emerged as a means to generate physics-grounded channel data for training such models \cite{nguyen2021digital, wu2023digital, li2024digital, he2023physics}. This work builds on both directions to compute accurate CSI under real-time constraints.
\subsection{Related Work}
Recent physical-layer research has increasingly adopted deep learning (DL) to alleviate the prohibitive overhead of CSI acquisition in massive MIMO and Wi-Fi systems \cite{guo2024deep, zhang2025predicting, swain2023low, srivastava2025deep}. To reduce the large data-collection burden of these models, several works turn to site-specific wireless digital twins and ray-tracing (RT) models \cite{alkhateeb2023real, luo2025digital, haider2025digital, alikhani2025digital, jiang2024digital}. By reconstructing the 3D geometry and the electromagnetic interactions of the environment, a learnable digital twin generates synthetic, high-fidelity CSI that closely resembles real-world propagation \cite{jiang2025learnable, tang2025semantic, haider2026llm, elloumi2025spectrum, wang2025radio, zhu2024toward, tarafder2026digital, iye2025open, oh2025digital, aram2025site}. Studies further show that fine-tuning DL-based CSI compression on these physics-inspired twins improves reconstruction quality compared to conventional schemes, provided the underlying geometry and hardware modeling preserve strict twinning fidelity \cite{luo2025digital, haider2025digital, jiang2024digital, saeizadeh2026airmap}. This trajectory moves from traditional statistical estimators toward data-driven frameworks that exploit simulated priors for real-time channel estimation. The CSI predicted by digital twins can be utilized for offline AI model training,  what-if analysis/network planning, and real-time proactive optimization, particularly in high-mobility scenarios where obtaining accurate field-level CSI is inherently challenging \cite{he2023physics, alkhateeb2023real, elloumi2025digital, makvandi2026adaptwin}.

In parallel, baseband processing has advanced through unified neural receivers that jointly perform multiple tasks. Recent frameworks replace disjoint signal-processing chains with multi-task DL that combines demodulation, channel decoding, and even source decoding within a single computational block \cite{sagduyu2024joint, wu2025low}, which reduces hardware complexity. This approach is effective across diverse physical layers, including long-haul optical links that pair a neural demodulator with soft-decision forward error correction \cite{mishina2025long}. To meet strict latency and storage budgets, other works apply tensor decomposition to compress these architectures without sacrificing link reliability \cite{wu2025low}.

Despite this progress, the following gaps remain. Existing digital twin channel models are rarely validated for generalization across distinct propagation environments, and they are seldom integrated with unified neural receivers to validate end-to-end performance of digital twin predicted channel under real-time latency constraints. Bridging these two directions requires an architecture that generalizes the channel twin across environments, unifies demodulation and decoding at the receiver, and reconstructs the full CSI from a sparse set of high-precision anchors within the microsecond budget of 6G numerologies. This work addresses that gap.

\subsection{Motivation \& Contribution}
\noindent\textbf{Motivation:}
High-fidelity CSI is fundamental for maximizing the spectral efficiency and the beamforming accuracy of massive MIMO, which is the core transmission technology for 6G. Channel twin represents an engine of two cascaded modules, such as site-specifc, calibrated ray-tracing engine and an AI model to correct the ray-tracing prediction as shown in \cite{haider2025digital}. However, acquiring precise channel knowledge in dynamic propagation environments remains a critical bottleneck. The generalizablity, and scalability of such channel twin \cite{haider2025digital} is not addressed, whereas its capability to operate in a fully pliot-less manner is not demonstrated. Calibrated 3D ray-tracing within a channel twin framework offers physics-grounded channel determinism, but its computational overhead makes per-symbol execution incompatible with the microsecond latency constraints of 5G NR and emerging 6G numerologies.   
For the 30~kHz subcarrier spacing, generating deterministic simulations for every OFDM symbol violates the real-time latency budget. This motivates a hybrid strategy that computes high-precision, physics-based CSI for a sparse subset of temporal anchor symbols and predicts the remaining symbols with a rapid interpolator.
\newline
The existing methods \cite{haider2025digital, luo2025digital, luo2026digital}, however, satisfy neither the accuracy nor the latency requirement of this sparse-anchor strategy. Traditional interpolators, such as splines, assume local channel smoothness and fail to capture the non-linear dynamics of multipath scattering and deep fades. Sequential models such as Long Short-Term Memory (LSTM) networks suffer from processing bottlenecks and localized receptive fields, and they struggle to bridge wide temporal gaps within the edge-inference latency limit. Therefore, a gap remains for a highly parallelizable, physics-aware architecture that leverages a global temporal receptive field to reconstruct the complete CSI matrix from sparse digital twin anchors under real-time constraints. The attention-based Transformer is well suited to this role.

\noindent\textbf{Contributions:}
The primary contributions of this work establish a comprehensive, AI-empowered Channel twin to reliably mimic the complex wireless propagation and predict highly accurate multi-path CSI for downstream applications such as precoding while avoiding costly pilot-based CSI acquisition for next-generation cellular networks. The contributions of this work can be distinguished by the following three novel advancements. 
\begin{itemize}
\item We introduce generalization of channel twin capability, a spatially transferable framework that leverages real-world field-collected CSI datasets from a mapped location to reliably infer channel characteristics in unseen environments, utilizing fine-tuned 3D ray-tracing models paired with AI-driven algorithms. This scene to scene transfer capability mitigates the need for exhaustive, site-specific empirical data collection.
\item The proposed framework abandons traditional, disjointed signal processing chains by integrating a unified neural receiver, replacing separate modular components by computing both demodulation and decoding tasks within a single AI-driven block to significantly reduce processing overhead, over-the-air pilot transmission and improve robustness against non-linear channel impairments.
\item To ensure the digital twin framework remains computationally tractable within the strict microsecond latency budgets of 5G NR numerologies, we scale the system by computing deterministic CSI for only a sparse subset of temporal anchors, utilizing the global receptive field of an attention-based Transformer model to dynamically and accurately predict the remaining symbols via parallelized inference.
\end{itemize}

This work substantially extends our earlier study in \cite{haider2025digital}, which introduced the U-Net based mapping from the RT-generated channel impulse response (CIR) to the real-world CIR for a single measured site. The present paper inherits this learned RT-to-real mapping as one building block and advances beyond it in three directions. First, the channel twin is generalized across spatially distinct environments and is adapted through transfer learning (TL), whereas \cite{haider2025digital} operated within a single scene. Second, the receiver is redesigned around a unified neural block that jointly performs the demodulation and the decoding. Third, the framework is made temporally scalable through the Transformer-based interpolation over sparse anchor symbols, together with the corresponding complexity analysis. None of these three components appears in \cite{haider2025digital}.

\section{System Model}

\begin{figure}[h!]
\centering
\includegraphics[width = 8cm, height = 6cm]{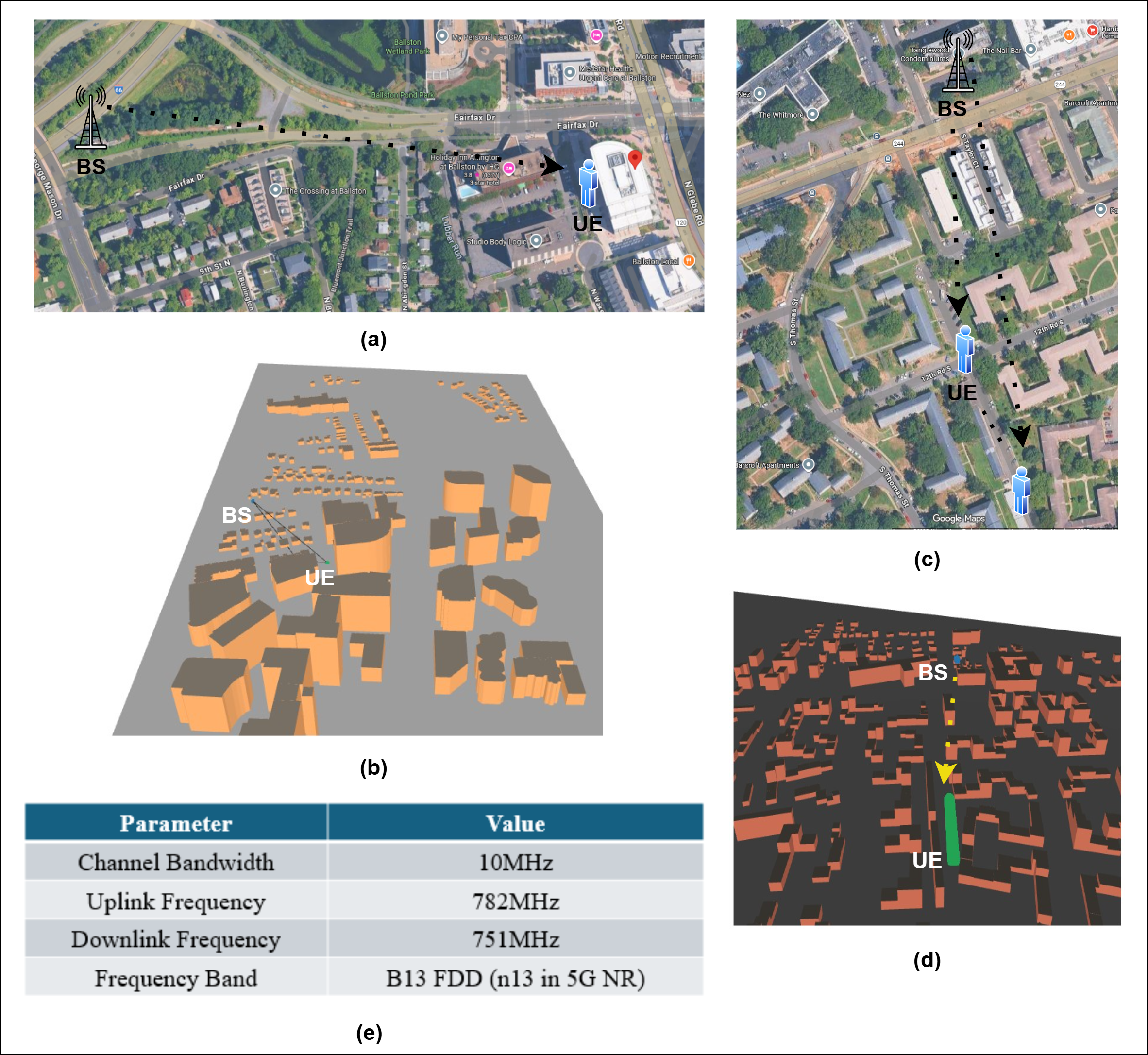}
\captionsetup{justification=centering}
\caption{Downlink MISO-OFDM communication system: (a) and (c) satellite views of Scene~1 and Scene~2, (b) and (d) the corresponding 3D ray-tracing models, and (e) the system parameters.}
\label{fig1}
\end{figure}

\noindent\textbf{Signal Model:} Fig.~\ref{fig1} shows a multiple-input single-output OFDM (MISO-OFDM) downlink communication system. We consider a time-varying frequency-selective multipath fading channel. Linear precoding is deployed to maximize the received signal power by exploiting transmit diversity. The base station (BS) is equipped with $N_{T}$ antennas, while the user equipment (UE) employs a single antenna for signal reception. The system operates over a transmission bandwidth divided into $N_{s}$ orthogonal subcarriers.
Let $s$ denote the transmitted information bits. At the BS, the information bits are first passed through a channel encoder that introduces controlled redundancy for protection against the channel impairments. We denote the channel encoding operation by $f(\cdot)$, and the resulting coded bits by $c = f(s)$. The coded bits are then mapped to baseband modulated symbols. The modulated symbol for the $k \in \{1,2,\ldots, N_{s}\}$-th OFDM subcarrier is denoted by $S_{k}$. These symbols are drawn from a standard constellation, such as binary phase-shift keying (BPSK), quadrature phase-shift keying (QPSK), or $M$-ary quadrature amplitude modulation ($M$-QAM). Let $B_{k}$ represent the precoding factor. It is a function of the underlying channel between the BS and the UE and depends on the adopted precoding scheme. The precoded signal for antenna $t \in \{1,2,\cdots,N_T\}$ is expressed as $X_{k}^t = S_{k}B_{k}^t$. Let $H_{k}^t$ denote the frequency-domain channel response for subcarrier $k$ over the link between antenna $t \in \{1,2,\cdots,N_T\}$ of the BS and the UE.
After the inverse fast Fourier transform (IFFT), the time-domain representations of $X_k^t$ and the CIR are denoted by $x_{n_1}^t[m]$ and $h_{n_2}^t[m]$, respectively, for OFDM symbol $m = \{1,2,\cdots\}$. Here, $n_1 \in \{1,2,\cdots,N_s\}$ and $n_2 \in \{1,2,\cdots,\eta_t\}$ are the indices of the time-domain signal and the CIR, respectively. The term $\eta_t$ denotes the length of the frequency-selective CIR for the link between antenna $t \in \{1,2,\cdots,N_T\}$ of the BS and the UE. For notational simplicity, we assume $\eta_t = \eta$ for all $t \in \{1,2,\cdots,N_T\}$. Adding a cyclic prefix (CP) of $\tau$ samples to each OFDM symbol $m$ yields $\Tilde{x}_n^t[m]$, with $n \in \{1,2,\cdots,N_s + \tau\}$.
The received signal vector $\yvec[m] \in \mathbb{C}^{1 \times (N_s + \tau + \eta - 1)}$ for OFDM symbol $m$ over the $N_T$ transmit antennas is expressed as
\begin{equation}
 \yvec[m]= \sum_{t=1}^{N_T} \Tilde{\xvec}^t[m] \circledast \hvec^t[m] + \wvec[m],
 \label{eq1}
\end{equation}
where $(\circledast)$ denotes the convolution operation. The vector $\wvec[m] \in \mathbb{C}^{1 \times (N_s + \tau + \eta - 1)}$ denotes the additive white Gaussian noise (AWGN), where each element has zero mean and variance $\sigma_w^2$. The terms $\Tilde{\xvec}^t[m] \in \mathbb{C}^{1 \times (N_s + \tau)}$ and $\hvec^t[m] \in \mathbb{C}^{1 \times \eta}$ denote the transmit signal vector and the CIR, respectively, for OFDM symbol $m \in \{1,2,\cdots\}$.
The UE removes the CP and performs the fast Fourier transform (FFT) to recover the frequency-domain signal. Two distinct channel estimates serve two distinct roles in this system. At the BS, the precoder is computed from the CSI inferred by the proposed channel twin, as detailed in Section~\ref{sec:framework}. At the UE, the channel used for data detection is obtained by pilot-based estimation, as depicted in Fig.~\ref{fig3}(a). The residual degradation caused by imperfect channel knowledge after the equalization stage is mitigated by the neural receiver introduced in Section~\ref{sec:nr}. In practice, blind estimation techniques can also be applied at the UE to estimate the CIR and thereby avoid pilot transmission, albeit at the cost of higher computational complexity. The transmit signal is then detected using linear or non-linear equalization. We denote the equalized signal by $\hat{S}_k$, which is demodulated to obtain the estimated coded bits $\hat{c}$. Finally, the channel decoder processes $\hat{c}$ to recover the estimated information bits $\hat{s}$. In the conventional receiver, the demodulation and the channel decoding are performed as two separate, sequential operations.

\noindent\textbf{Precoding Schemes:}We consider a linear precoding scheme, namely minimum mean square error (MMSE). 
For MMSE precoding, the precoding factor is given by $B_k^t = (\hat{H}_k^t)^{*} (\hat{H}_k^t (\hat{H}_k^t)^{*} + \sigma_w^2 / {P_t})^{-1}$ \cite{selvan2014performance}, where $(\ast)$ denotes the conjugate operation, $\hat{H}_k^t$ is the estimated channel frequency response, and $P_t$ is the transmit power for antenna $t \in \{1,2,\cdots,N_T\}$. It is worth noting that the objective of this paper is to obtain $\hat{H}_k^t$ precisely from the environment-aware channel twin model.

\section{Proposed Framework}\label{sec:framework}

\noindent The proposed channel twin framework consists of three components: the generalization of the channel twin for spatial scalability, a unified neural receiver for joint demodulation and decoding for performance enhancement, and a Transformer-based interpolator for temporal scalability. We detail each component in the following.
\subsection{Generalization of Channel Twin}
\begin{figure*}
    \centering
    \includegraphics[width=0.95\linewidth]{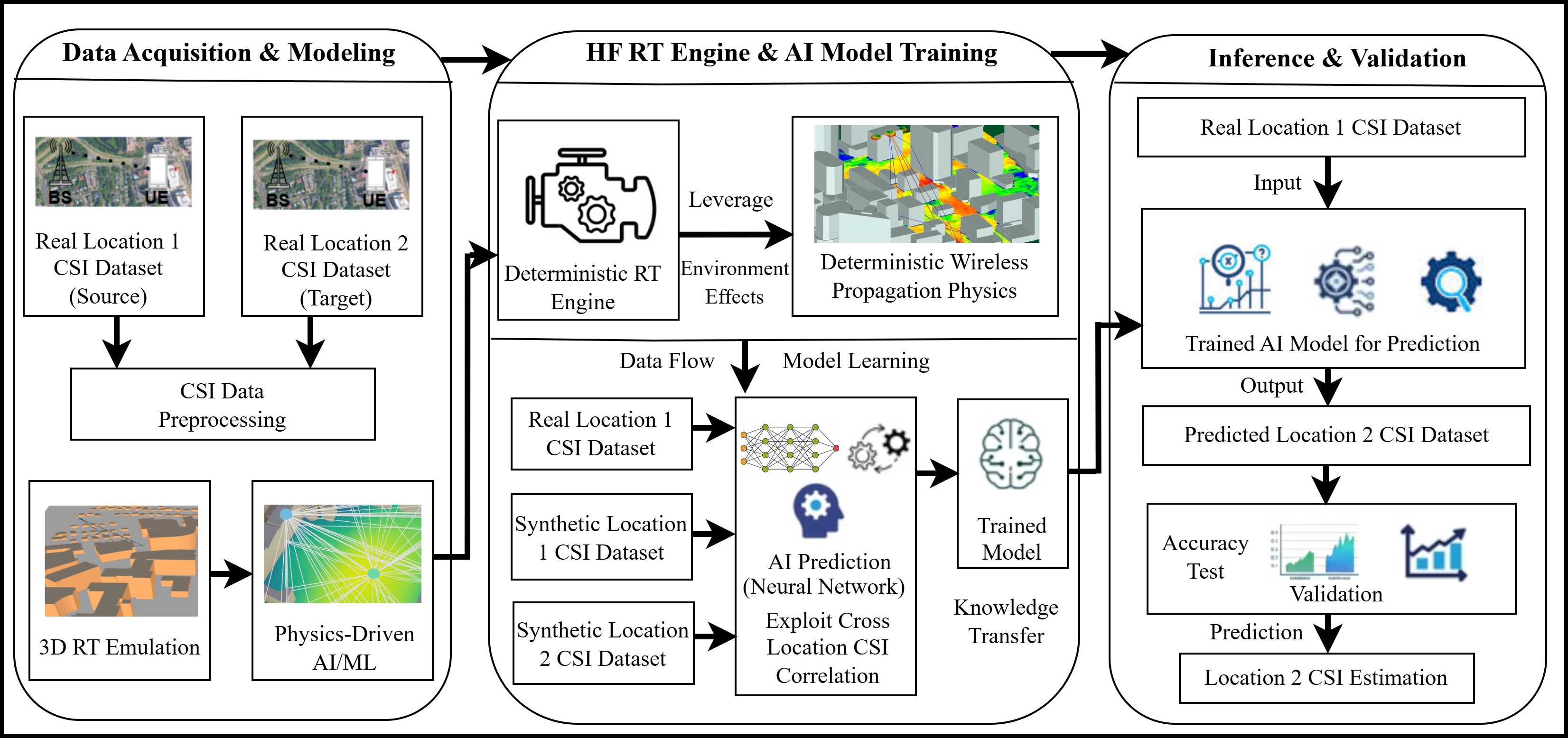}
    \caption{Generalization of channel twin.}
    \label{fig2}
\end{figure*} 
It is worth noting here that the term channel twin denotes the propagation-specific component of the wireless digital twin, namely the calibrated 3D RT model together with the AI model that maps its output close to the counterpart real-world underlying wireless channel. We adopt the process of designing reliable and fine tuned channel twin from \cite{haider2025digital}. A generalized wireless channel twin is a high-fidelity, dynamic virtual replica of the physical radio propagation environment. It is a key enabler of the predictive network capabilities required by 6G and advanced MIMO systems. Its main role is to shift network operation from reactive troubleshooting to proactive optimization, which supports seamless resource allocation, dynamic network optimization, and robust multipath mitigation. A reliable twin fuses deterministic modeling, such as 3D RT, with continuous empirical measurements, since the physical environment is highly non-stationary. High-precision calibration can be achieved through a closed-loop feedback mechanism that minimizes the divergence between the digital simulation and the real-world measurements. In this paper, the calibration is performed offline. The radio material properties and the key RT parameters are tuned once against the drive-test measurements, as described in Section~\ref{sec:results}, and the continuous closed-loop refinement is left as a deployment capability beyond the present scope.

\noindent Fig.~\ref{fig1}(a) and Fig.~\ref{fig1}(c) show two nearby scenes in Northern Virginia, USA, and Fig.~\ref{fig1}(b) and Fig.~\ref{fig1}(d) show the corresponding 3D RT models. The system parameters are summarized in Fig.~\ref{fig1}(e). We captured the demodulation reference signal from the nearby BS using a software-defined radio through a drive test. Scene~1 lies along North Glebe Road, Ballston, Arlington, at latitude 38.88175\degree~N and longitude 77.12180\degree~W, and Scene~2 lies along South Taylor Street, Columbia Pike, Arlington, at latitude 38.85615\degree~N and longitude 77.10294\degree~W. The prediction of the CSI of an unobserved target location from the CSI of a known reference location relies on the spatial correlation of the electromagnetic field and the shared geometry of the environment. Intuitively, nearby locations are illuminated by the same macroscopic scattering clusters. Therefore, their channel structures, although distinct, share coupled propagation paths and delay profiles. The generalization studied in this paper is therefore scoped to nearby environments with shared morphology, and the two measured scenes in Fig.~\ref{fig1} satisfy this condition. To exploit this property, an AI-driven model learns the latent mapping that transforms the multipath profile of the observed anchor location into the localized fading characteristics of the target location. As a result, the channel twin extrapolates the CSI across a spatial grid. This capability reduces the need for dense pilot signaling and enables real-time kinematic positioning and optimal beam selection with minimal overhead.

\noindent Fig.~\ref{fig2} presents a high-level overview of the generalization process. The objective is to predict the CSI of a target location from the CSI of a known location, using RT and an AI-driven algorithm that captures the correlation between the two locations from the environmental physics and the wireless propagation effects. As shown in Fig.~\ref{fig2}, the process is organized into three functional phases.

\noindent \textbf{Data Acquisition \& Modeling:} This phase captures the initial real-world measurements and establishes the environmental context. It preprocesses the source scene dataset and the target scene ground-truth dataset, and it builds the 3D RT models required for the simulation stage.

\noindent \textbf{Ray-Tracing \& AI Model Training:} This is the core engine of the framework. The RT engine uses the 3D models to generate synthetic datasets that capture the site-specific physics of signal propagation. The AI model \cite{haider2025digital} is then trained on both the real CSI and these physics-grounded synthetic datasets from the fine-tuned RT model to learn the correlation between the two scenes.

\noindent \textbf{Inference \& Validation:} Once trained, the model enters the operational phase. Live data from the source scene is input to the model to generate a CSI prediction for the target scene. This phase includes a validation step that compares the prediction against the real ground truth to verify the estimation accuracy.

\noindent This flow integrates the calibrated RT 3D modeling with data-driven AI for accurate site-specific CSI estimation, which supports efficient wireless network planning and optimization. We evaluate the precision of the target-scene CSI estimation over three test cases, together with a scalability study of the temporal interpolator. The corresponding numerical results are presented in Section~\ref{sec:results}.
\subsection{Performance Enhancement Using Neural Receiver}\label{sec:nr}
\begin{figure*}
    \centering
    \includegraphics[width=0.95\linewidth]{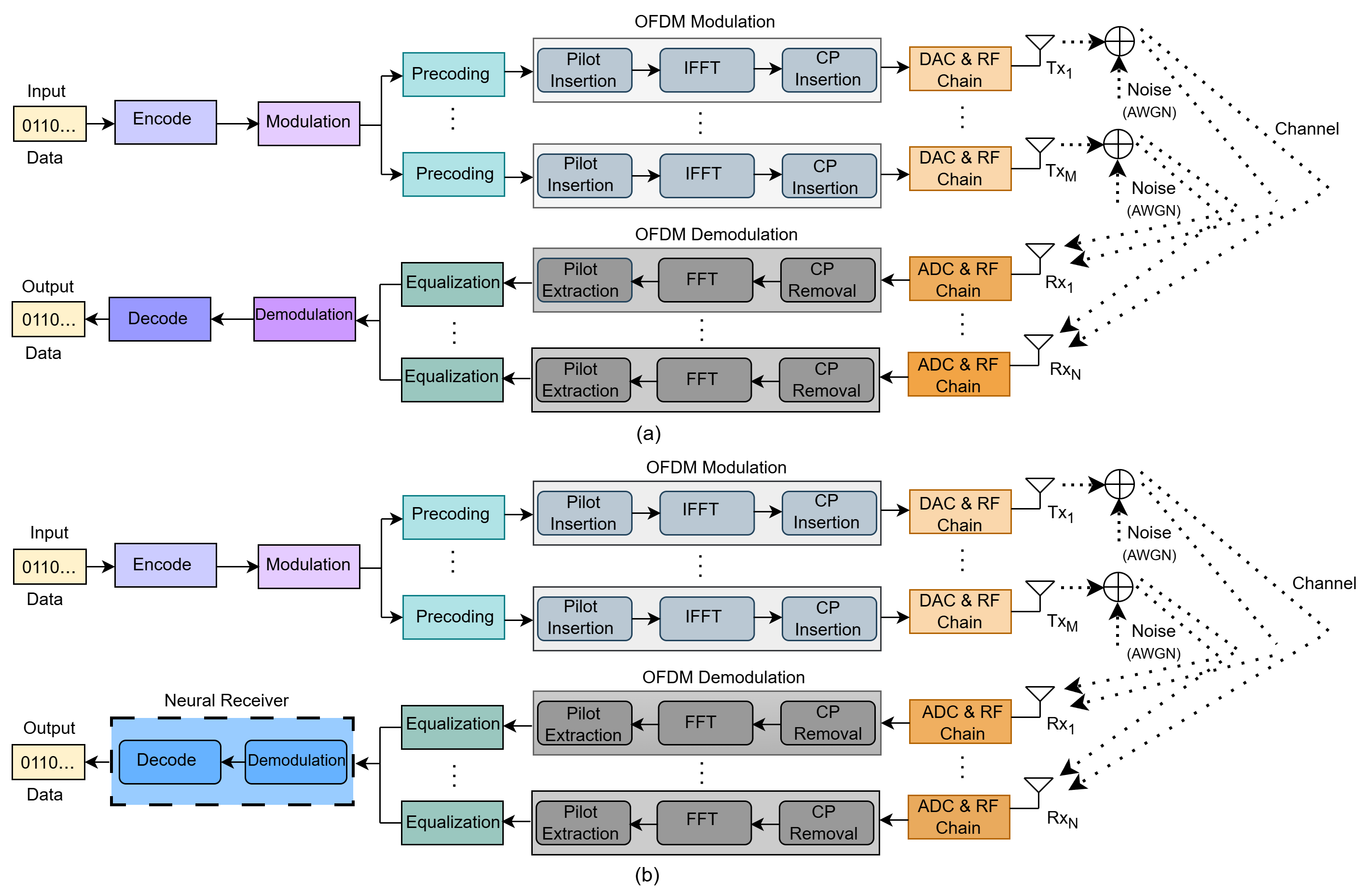}
    \caption{Block diagram of end-to-end communication system.}
    \label{fig3}
\end{figure*} 
Fig.~\ref{fig3}(a) shows a conventional end-to-end MISO-OFDM receiver that employs QPSK modulation and Hamming channel coding. Following the transmission of the Hamming-encoded, QPSK-modulated symbols over the frequency-selective multipath channel, the conventional receiver applies a sequence of disjoint, analytically derived processing blocks: OFDM demodulation, pilot-based channel estimation, linear equalization such as zero-forcing (ZF) or MMSE, and demodulation to generate the log-likelihood ratios (LLRs) for the Hamming decoder. Although this modular pipeline is computationally well understood, it is fundamentally limited by its assumption of a linear, Gaussian channel and perfect synchronization. In practical environments characterized by severe multipath fading, inter-antenna interference, and hardware non-linearities, the rigid separation of channel estimation and equalization causes compounded error propagation. As a result, the reliability of the soft bits fed to the Hamming decoder degrades, which increases the overall bit error rate (BER).


\noindent In Fig.~\ref{fig3}(b), the conventional demodulation and decoding blocks are replaced by a unified, data-driven neural receiver. The front-end processing remains conventional. The receiver performs the OFDM demodulation, the pilot-based channel estimation, and the linear equalization exactly as in Fig.~\ref{fig3}(a). The neural receiver then takes the equalized symbol grid $\hat{S}_k$ as its input and jointly performs the demodulation and the channel decoding within a single computational block. Its output is the estimated information bits $\hat{s}$. Trained on diverse channel realizations, the network learns the soft mapping from the equalized symbols to the information bits. Hence, it avoids the hard, sequential hand-off between a separate demapper and a separate Hamming decoder. Moreover, it merges the separate demodulation and decoding blocks into a single block. Hence, it improves the energy efficiency and reduces the hardware complexity at the receiver.


\noindent For the evolution toward 6G cellular networks, the neural receiver becomes a structural necessity rather than a mere performance optimization. Classical demodulation and decoding algorithms rely on tractable, stationary mathematical models that degrade rapidly under the heterogeneous conditions of 6G networks. The neural receiver instead adopts a data-driven function approximator. By treating the demodulation and the decoding as a single learning task, the network learns to map the equalized symbols directly to the information bits. It therefore exploits the code structure and the constellation geometry jointly, and it compensates for residual impairments that are difficult to model in closed form after the equalization stage.


\noindent \textbf{Neural Receiver Architecture:} The neural receiver is a fully connected feed-forward network that consists of an input layer, $\mathcal{L} - 2$ hidden layers, and a final output classification layer, where $\mathcal{L}$ denotes the total number of layers. The input layer receives the equalized symbol grid $\hat{S}_k$, and the output layer produces the estimated information bits $\hat{s}$. Since the receiver is configured at the UE, where the computational resources are limited, a lightweight fully connected feed-forward model is adopted in place of a high-complexity neural network model. Each hidden layer contains $Q$ neurons.

\subsection{Scalability of Channel Twin}
\begin{figure}[h!]
\centering
\includegraphics[width = 8cm, height = 3cm]{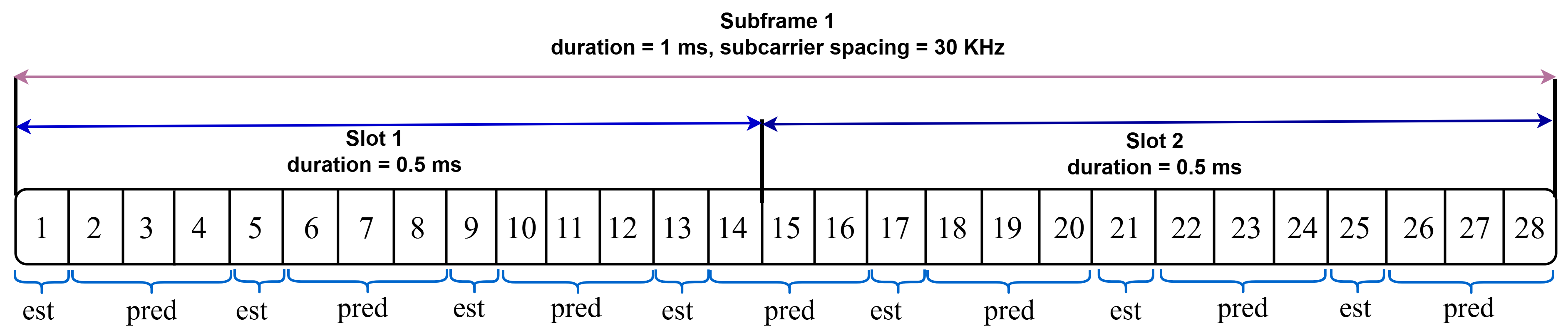}
\captionsetup{justification=centering}
\caption{Time slot specification for interpolation.}
\label{fig4}
\end{figure}
The integration of 3D digital twins and AI-driven RT offers high accuracy for CSI acquisition, but the associated computational load introduces a critical latency bottleneck. For the 5G NR 30~kHz numerology, computing these precise, physics-based CSI values for every OFDM symbol is prohibitive within the strict microsecond latency budget. To resolve this, we adopt a selective computation strategy. As shown in Fig.~\ref{fig4}, the 30~kHz numerology contains a total of 28 OFDM symbols per subframe, organized into two slots of 14 symbols each. A designated subset of these symbols serves as the estimation symbols, also referred to as the anchor symbols or the temporal anchors throughout this paper, for which the CSI is computed deterministically by the RT engine and fine tuned by the AI-driven algorithm \cite{haider2025digital}. RT engine identifies dominant multi-path components (MPCs), including LoS, reflection, scattering, and refraction and returns their complex path gains, and delay profiles. Using these information into  $h_{rt}(\tau) = \sum_{k=1}^{N}a_{k}e^{-j2\pi f\tau_{k}}\delta(\tau-\tau_{k})$, where $f$, $k\in\{1,2,..., N\}$, $a_{k}$, $\tau_{k}$, and $\delta(\tau)$ represent operating frequency, the number of paths, the amplitude of path $k$, the time delay of path $k$, and Dirac delta function of the impulse response of the multipath component, respectively \cite{haider2025digital}. The CSI of the remaining symbols is then obtained by AI-enhanced interpolation.
This hybrid strategy reduces the processing overhead and the power consumption without sacrificing the spatial intelligence provided by the digital twin, which ensures near-instantaneous, high-fidelity channel estimation for latency-critical applications.

\noindent The feasibility of this interpolation framework stems from the high temporal correlation of wireless channels at millisecond timescales. Since the propagation environment changes smoothly across adjacent OFDM symbols, the AI model accurately extrapolates the missing CSI from the sparse, high-precision anchor points generated by the RT engine. This assumption is quantified as follows. At the carrier frequency of 751~MHz, a vehicular speed of up to 100~km/h yields a maximum Doppler shift of approximately 70~Hz and a channel coherence time of approximately 6~ms. Hence, the coherence time exceeds the 1~ms subframe duration by a wide margin, and the channel remains highly correlated across the 28 symbols spanned by the anchors. The CSI of the estimation symbols is generated dynamically by the fine-tuned RT 3D model and the AI-driven algorithm, which leverage site-specific wireless propagation physics. By learning the spatial and temporal correlation of the CSI at the estimation symbols, the model accurately predicts the CIR of the remaining symbols.

\noindent \textbf{Scalable Interpolator Architecture:} We adopt a Transformer neural network to design the scalable interpolator. The core of the Transformer is a stack of identical encoder and decoder layers. Each encoder layer contains two primary sub-layers, namely a multi-head self-attention mechanism and a position-wise feed-forward network. First, an embedding layer maps each token, which in this case is a CSI data point, into a vector space, and positional encodings are added to retain the sequence order. The self-attention mechanism then computes the context weight of each token with respect to every other symbol in the subframe, and it processes all tokens in parallel. Multi-head attention extends this by executing several attention operations simultaneously and merging their outputs. A residual connection and layer normalization are applied around each sub-layer, and the feed-forward network provides an additional non-linear mapping. This parallelized structure enables the encoder to build a globally contextualized representation of the subframe from the sparse anchors.

\noindent The decoder reconstructs the sequence using a modified stack of layers that introduces a third sub-layer, namely cross-attention. During training, a masked multi-head self-attention mechanism prevents the decoder from attending to future positions. Since the interpolation task provides all the anchor symbols of the subframe in advance, this causality constraint is relaxed in our implementation, and the decoder reconstructs all the symbol positions jointly in a single parallel pass. The cross-attention mechanism then takes the encoded anchor context and computes attention weights relative to the current decoder state. This allows the decoder to query the global environmental knowledge while generating the complete sequence of 28 high-resolution CSI matrices of the subframe. Residual connections, layer normalization, and a position-wise feed-forward network are applied at each sub-layer. Finally, a linear layer projects the decoder output back to the CSI dimension and reconstructs the entire subframe.

\subsection{Complexity Analysis and Real-Time Feasibility}\label{sec:complexity}
We now quantify the computational cost of the proposed scheme during the data transmission phase and justify its real-time scalability. Each subframe of the 30~kHz numerology contains $N_{\mathrm{sym}} = 28$ OFDM symbols. The CSI of $N_{\mathrm{est}} = 7$ estimation symbols is computed by the RT engine and the trained U-Net, and the CSI of the remaining $21$ symbols is predicted by the trained Transformer. Both networks are trained offline. Hence, only the inference complexity affects the real-time budget of one subframe.

\noindent \textbf{Estimation Complexity:} The cost of one RT execution scales as $\mathcal{O}(N_{\mathrm{ray}}\,\beta)$, where $N_{\mathrm{ray}}$ is the number of projecting rays and $\beta$ is the maximum interaction order of reflections, scattering and diffractions. We deliberately project a small number of rays at the estimation symbols. Therefore, the per-symbol RT cost remains low. The resulting low-fidelity CIR is refined by the trained U-Net, which maps it to a close-to-real-world CIR. Since the U-Net is a convolutional encoder-decoder with a fixed number of layers $L_{\mathrm{u}}$, kernel size $\kappa$, and channel width $C$, its inference cost is $\mathcal{O}(L_{\mathrm{u}}\,\kappa\,C^{2}\,\eta)$, which is linear in the CIR length $\eta$. Hence, the total estimation cost per subframe is $\mathcal{O}\!\left(N_{\mathrm{est}}\,(N_{\mathrm{ray}}\,\beta + L_{\mathrm{u}}\,\kappa\,C^{2}\,\eta)\right)$. The fidelity lost by reducing $N_{\mathrm{ray}}$ is recovered by the learned RT-to-real mapping rather than by additional ray projection. Therefore, the accuracy burden is shifted from the RT engine to a fixed-cost neural inference.

\noindent \textbf{Prediction Complexity:} The Transformer interpolator processes the subframe as a sequence of $N_{\mathrm{sym}}$ tokens with model dimension $d_{\mathrm{m}}$ and $L_{\mathrm{tr}}$ layers. Its inference cost is $\mathcal{O}\!\left(L_{\mathrm{tr}}\,(N_{\mathrm{sym}}^{2}\,d_{\mathrm{m}} + N_{\mathrm{sym}}\,d_{\mathrm{m}}^{2})\right)$, where the first term accounts for self-attention and the second term accounts for the position-wise feed-forward layers. Since $N_{\mathrm{sym}} = 28$ is small and fixed by the numerology, the quadratic term is modest. More importantly, the entire subframe is processed in a single parallel pass with a sequential depth of $\mathcal{O}(1)$. In contrast, an LSTM requires $\mathcal{O}(N_{\mathrm{sym}}\,d_{\mathrm{m}}^{2})$ operations with a sequential depth of $\mathcal{O}(N_{\mathrm{sym}})$, and a bidirectional LSTM doubles this sequential depth. Hence, the Transformer attains a lower wall-clock latency on parallel hardware although its operation count contains a quadratic term.

\noindent \textbf{Training Complexity:} The offline training cost of each network is a constant multiple of its forward-pass cost per sample, accumulated over the dataset and the training epochs. For the U-Net, the training cost is $\mathcal{O}(E_{\mathrm{u}}\,D_{\mathrm{u}}\,L_{\mathrm{u}}\,\kappa\,C^{2}\,\eta)$, where $E_{\mathrm{u}}$ and $D_{\mathrm{u}}$ denote the number of epochs and the number of training samples, respectively. For the Transformer, the training cost is $\mathcal{O}\!\left(E_{\mathrm{tr}}\,D_{\mathrm{tr}}\,L_{\mathrm{tr}}\,(N_{\mathrm{sym}}^{2}\,d_{\mathrm{m}} + N_{\mathrm{sym}}\,d_{\mathrm{m}}^{2})\right)$. These costs are incurred once, before deployment. Therefore, they do not affect the per-subframe latency budget.

\noindent \textbf{Scalability Rationale:} The proposed scheme is scalable for three reasons. First, the deterministic RT computation is confined to $N_{\mathrm{est}}/N_{\mathrm{sym}} = 25\%$ of the symbols, which reduces the RT invocation rate by a factor of four. Second, the per-invocation RT cost is tunable through $N_{\mathrm{ray}}$, and the trained U-Net compensates for the reduced ray count. Hence, the estimation cost is decoupled from the geometric complexity of the environment. Third, the prediction cost of the Transformer is independent of the propagation environment and is executed as a single parallel inference per subframe. Consequently, the end-to-end CSI acquisition for one subframe consists of seven low-cost RT executions, seven U-Net inferences, and one Transformer inference, all of which fit within the 1~ms subframe duration. This combination of a reduced RT duty factor, a tunable ray budget, and a constant-depth interpolator constitutes the basis of our real-time scalability claim.

\section{Simulation Results}\label{sec:results}

\subsection{Parameter Specifications}
In this section, we present the numerical performance evaluation of the proposed scheme, which produces a CIR that closely matches the real-world CIR. We adopt the BER as the principal performance metric. The BER results are reported against the effective signal-to-noise ratio (SNR), defined as $\gamma \triangleq 1/\sigma_{w}^{2}$, where $\sigma_{w}^{2}$ denotes the noise variance. The quantity $\gamma$ serves as the abscissa of all the curves reported in this section. The evaluation is conducted over a live 751~MHz cellular deployment, which provides an accessible and repeatable measurement testbed. The proposed methodology, namely the RT-based anchor generation, the learned RT-to-real mapping, and the Transformer-based interpolation, is agnostic to the carrier frequency. 

We further demonstrate how the SE improves as the received SNR is maximized through precise beamforming. The received SNR $\gamma_R$ at the UE is expressed as
\begin{equation}
\gamma_R = \frac{P_{tx}\,\lvert \mathbf{h}_{actual}^{H}\mathbf{w}\rvert^{2}}{\sigma_{w}^{2}} = \frac{P_{tx}\,g}{\sigma_{w}^{2}},
\label{eq:rxsnr}
\end{equation}
where $P_{tx}$ is the transmit power, $\mathbf{h}_{actual}$ is the actual propagation channel of the environment, $\mathbf{w}$ is the precoding weight vector, and $g = \lvert \mathbf{h}_{actual}^{H}\mathbf{w}\rvert^{2}$ denotes the effective beamforming gain. The precoder $\mathbf{w}$ is designed from the estimated channel $\hat{\mathbf{h}}$. Therefore, the received SNR is governed by the alignment between the precoder and the actual channel, which in turn depends on the accuracy of $\hat{\mathbf{h}}$. The two quantities $\gamma$ and $\gamma_R$ coincide under unit transmit power and unit effective beamforming gain.

The computation of $g$ is consistent with the CSI-based beamforming approach adopted in practical cellular network standards. In practical frequency-division duplex (FDD) MIMO systems, the BS transmits pilot sequences. The user estimates the CSI from these pilots and feeds it back to the BS. The BS then exploits this CSI to compute the optimal precoding vector. However, this procedure incurs significant signaling overhead and is highly susceptible to pilot contamination, particularly when the BS employs large-scale antenna arrays, as expected in the 6G deployments in frequency range 2 and frequency range 3 (FR2/FR3). To address these limitations, we obtain the precise CSI estimate $\hat{\mathbf{h}}$ from the proposed environment-aware channel twin. The channel twin anticipates a rich set of site-specific context, including geospatial features, BS antenna characteristics, the radio material properties of 3D objects, the user location, and high-fidelity RT-enabled RF propagation with multi-interaction effects. The resulting digital twin generated CSI is then used to select the precoding vector and the beamforming gain that maximize the received SNR. This mitigates the pilot overhead and the pilot contamination inherent in conventional FDD MIMO systems \cite{alkhateeb2023real}.

The primary objective is to maximize $g$, which in turn maximizes the received SNR. A higher precision in estimating the underlying CSI yields a more accurate precoder, and therefore a higher achievable SNR. The SE follows as $\mathrm{SE} = \log_{2}(1 + \gamma_R)$. The reported SE is the Shannon spectral efficiency evaluated at the achieved received SNR. It represents the rate attainable when an adaptive modulation and coding scheme matches the transmission rate to $\gamma_R$, as adopted in practical cellular standards. Hence, the SE curves quantify the beamforming quality enabled by each channel estimation scheme. A more accurate CSI estimate yields a better aligned precoder, a larger effective beamforming gain, and therefore a higher attainable rate. The absolute values of the reported SE depend on the scaling convention of the underlying CIR datasets. The RT-generated and the real-world CIR datasets are scaled by a common factor before processing. Hence, the effective beamforming gain $g$ absorbs this common scale, and the received SNR in \eqref{eq:rxsnr} is defined with respect to the scaled channel rather than a channel normalized to the antenna count. All the schemes in this paper are evaluated under the identical scale. Therefore, the SE comparisons across the schemes are scale-invariant, and the relative gaps among the curves quantify the precoder alignment enabled by each channel estimation method.

A precise computation of the channel inversion is assumed throughout the precoding stage, and the precoder is obtained using the MMSE scheme of Section~II.
We adopt the standard Hamming code $(n, k)$ with $n = 7$ and $k = 4$ for channel encoding and decoding. This short code is chosen deliberately as a lightweight evaluation link. The integration of the standardized low-density parity-check (LDPC) and polar codes into the proposed framework is left for future work. For data detection, we employ the QPSK modulation scheme together with a ZF equalizer. For the CIR prediction task, the scaling of the RT-generated and the real-world CIR datasets, the partitioning of the data into training and testing subsets, and the deep neural network modeling are carried out using the MATLAB Deep Learning Toolbox through Monte Carlo simulations. Unless otherwise stated, we set $N_{T} = 4$ transmit antennas at the BS and a single receive antenna at the UE, $N_{s} = 128$, $\tau = 16$, and $\mathcal{L} = 5$. We adopt the U-Net model to obtain precise CSI from the tuned high-fidelity digital twin (HF-DT) dataset \cite{haider2025digital}. The learning rate is set to $\xi = 0.0008$.
After importing the RT emulation-generated CIR into MATLAB, we first scale both the RT-generated and the real-world CIR datasets. The sampling frequency is set to $\mathcal{F}_{s} = 30.72$~MHz. The measurement and dataset details are as follows. The drive test was conducted at a UE speed of approximately 25~mi/h, and the demodulation reference signal was captured over the B13 downlink at 751~MHz. The captured dataset comprises $10^{5}$ CIR snapshots for Scene~1 and $10^{5}$ CIR snapshots for Scene~2. For each scene, the dataset is partitioned into 80\% for training and 20\% for testing.

\subsection{Generalization of Channel Twin}
\begin{figure}[h!]
\centering
\includegraphics[width = 8cm, height = 6cm]{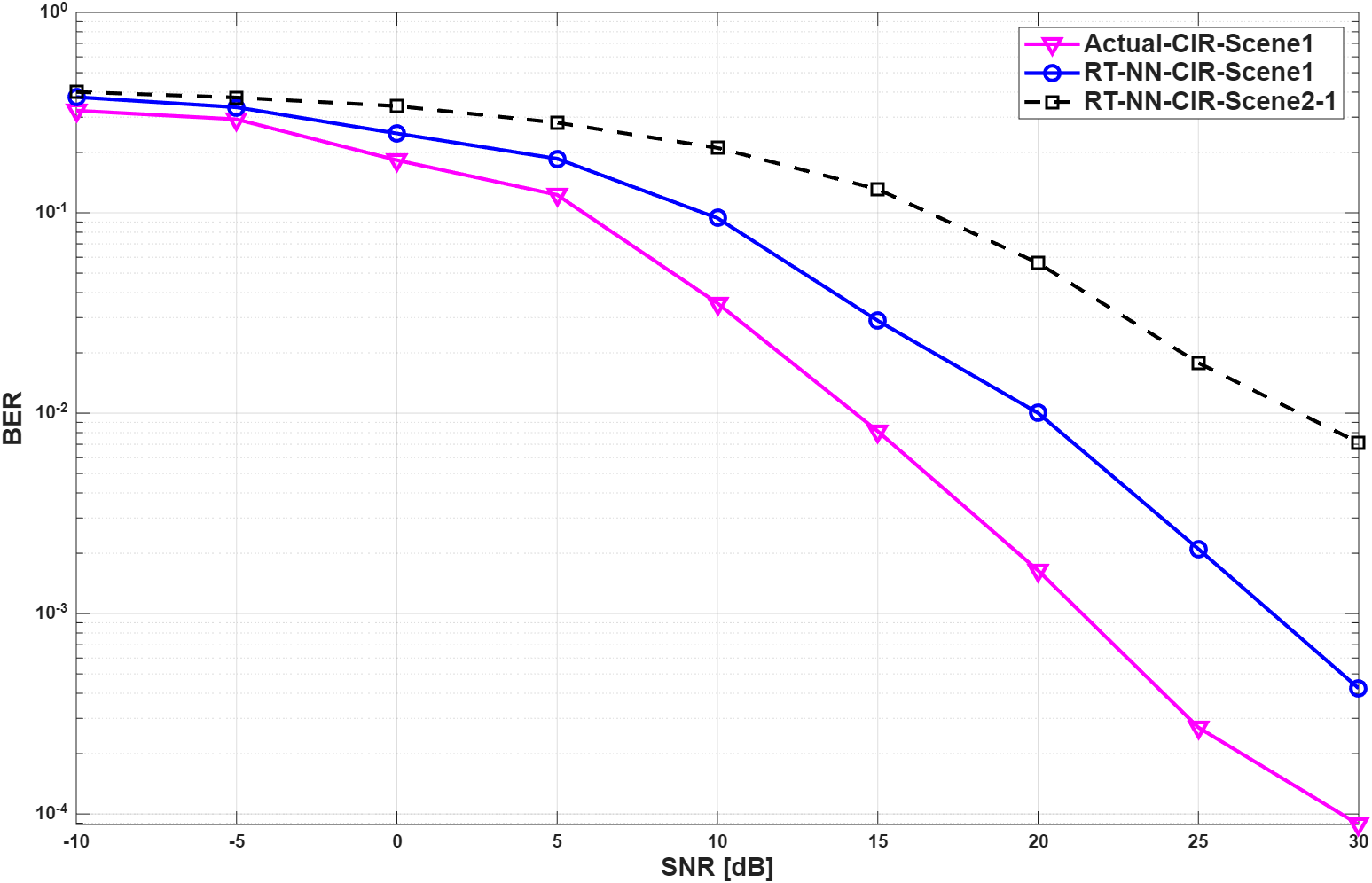}
\captionsetup{justification=centering}
\caption{BER versus SNR curve for test case 1.}
\label{fig5}
\end{figure}

\begin{figure}[h!]
\centering
\includegraphics[width = 8cm, height = 6cm]{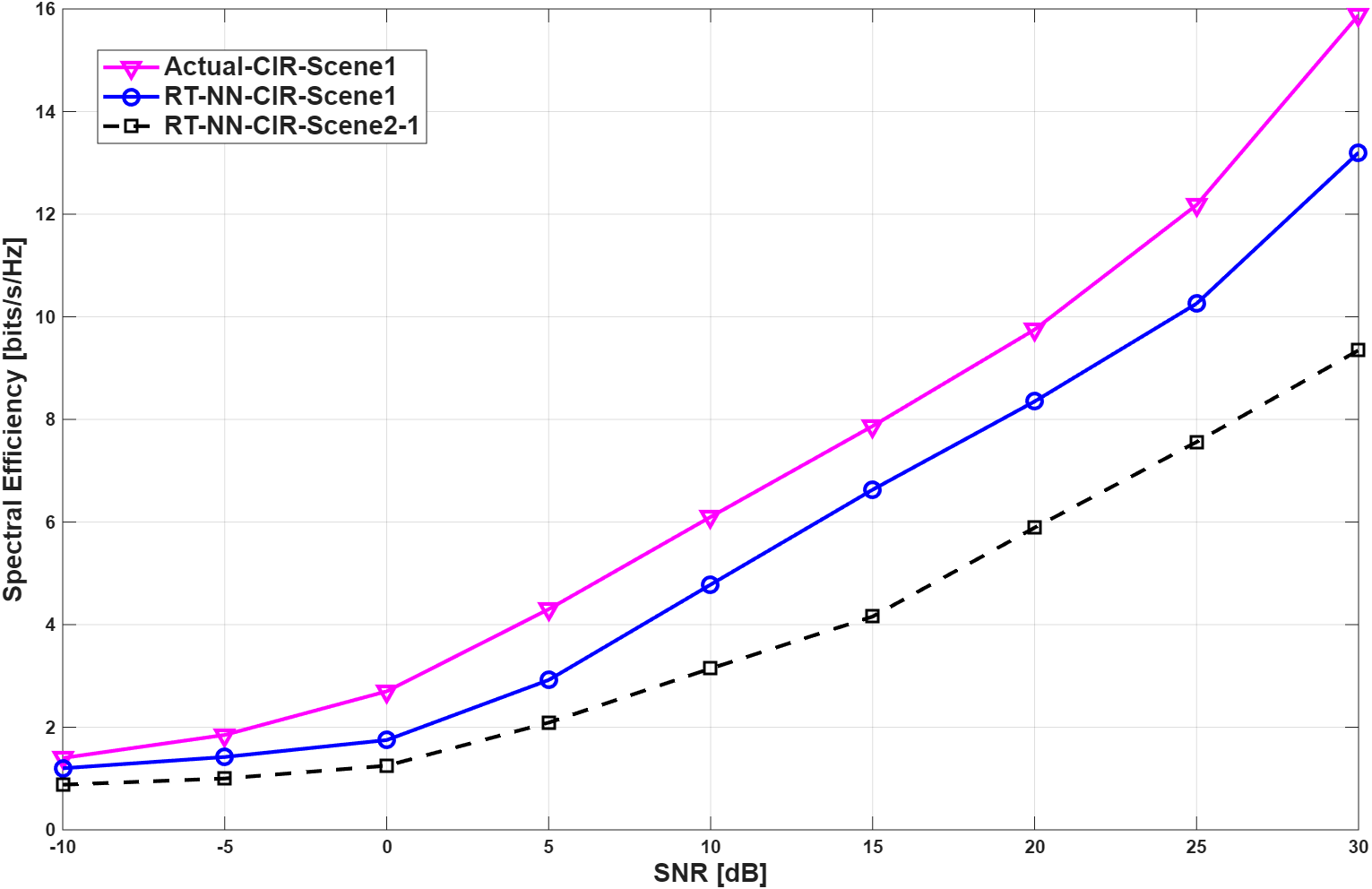}
\captionsetup{justification=centering}
\caption{Spectral efficiency versus SNR curve for test case 1.}
\label{fig6}
\end{figure}

\begin{figure}[h!]
\centering
\includegraphics[width = 8cm, height = 6cm]{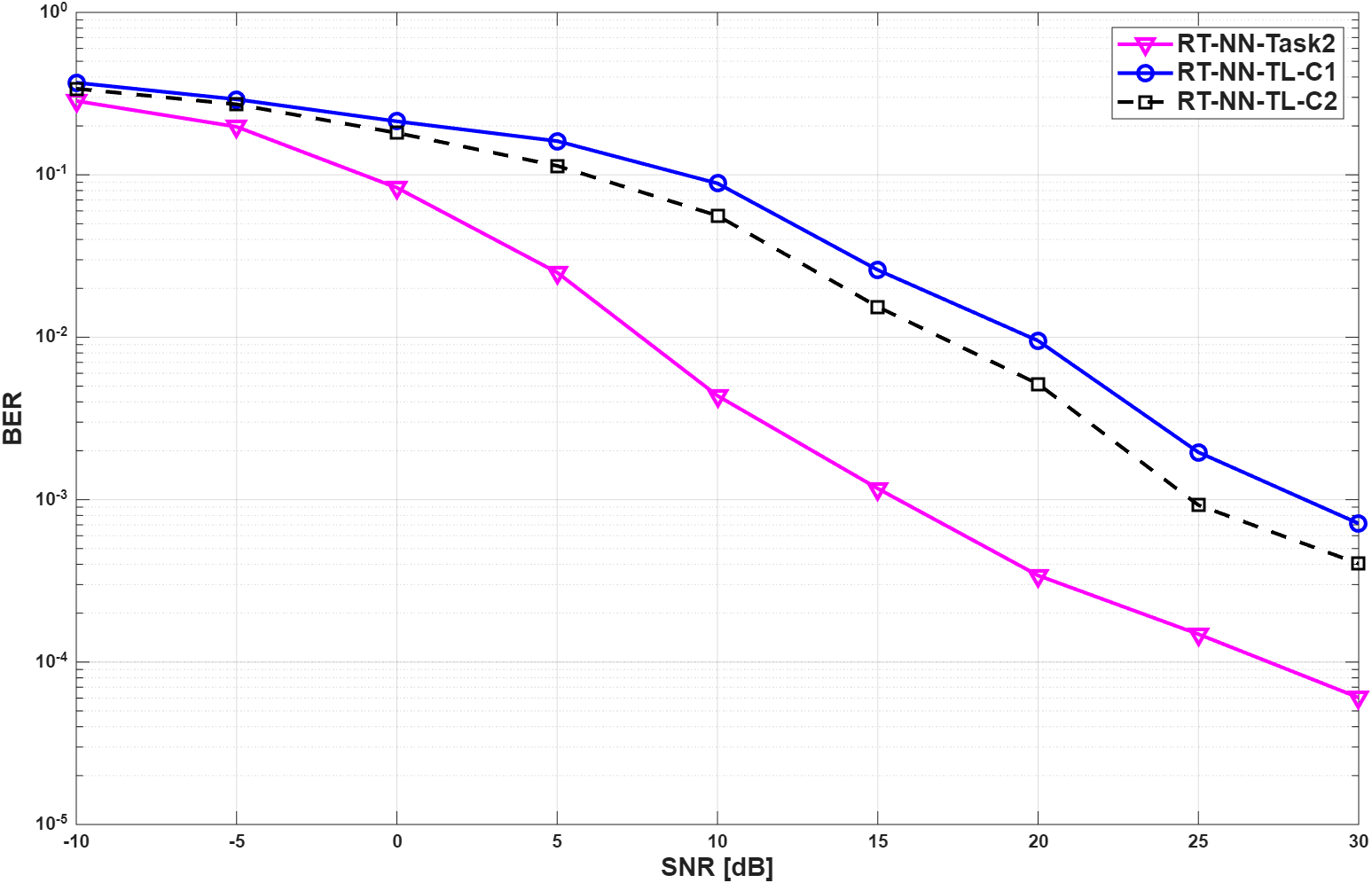}
\captionsetup{justification=centering}
\caption{BER versus SNR curve for test case 2.}
\label{fig7}
\end{figure}

\begin{figure}[h!]
\centering
\includegraphics[width = 8cm, height = 6cm]{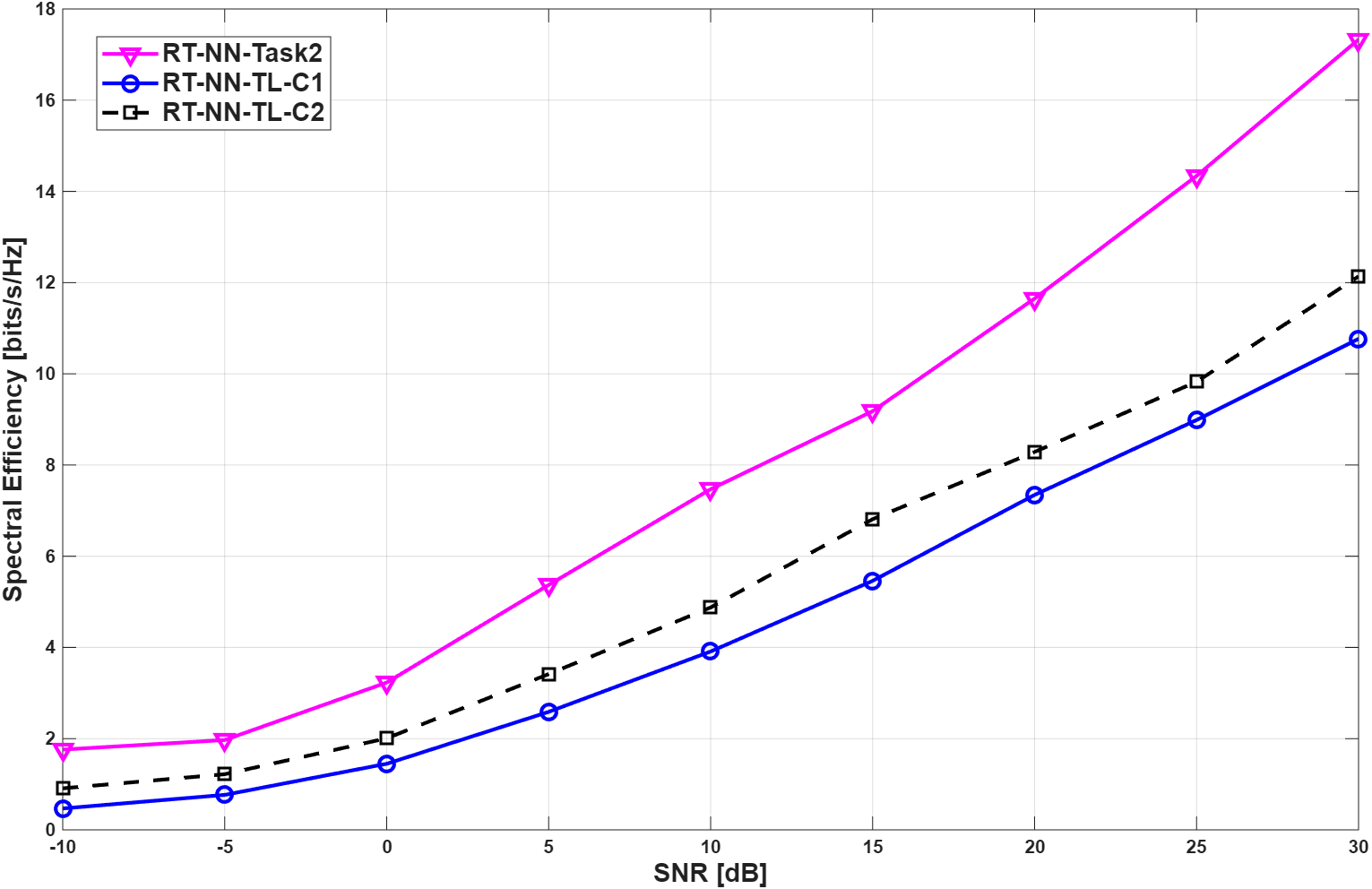}
\captionsetup{justification=centering}
\caption{Spectral efficiency versus SNR curve for test case 2.}
\label{fig8}
\end{figure}

\begin{figure}[h!]
\centering
\includegraphics[width = 8cm, height = 6cm]{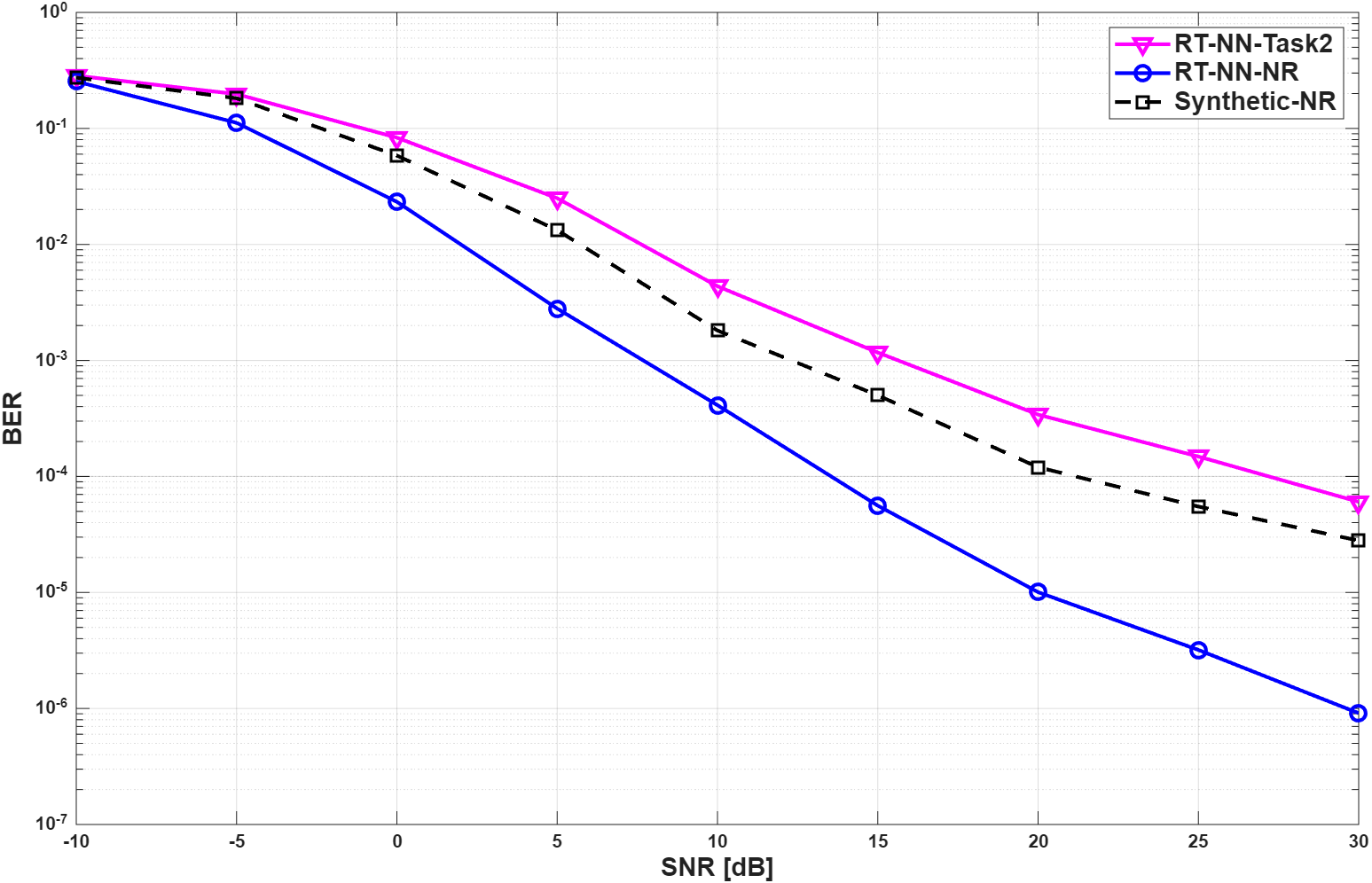}
\captionsetup{justification=centering}
\caption{BER versus SNR curve for test case 3.}
\label{fig9}
\end{figure}
In Fig.~\ref{fig5}, we evaluate the spatial transferability of the proposed HF-DT framework. In this framework, a U-Net neural network (NN) learns the correlation between the RT-generated CIR and the real-world CIR captured during the drive test. We refer to the proposed scheme as the HF-DT method, since the material properties and the key RT parameters, such as the number of projecting rays, are optimized. This optimization makes the synthetic CIR closely resemble the real-world measurements. The two scenes are the drive test sites described in Section~\ref{sec:framework}. For this test case, all the schemes are evaluated on Scene~1, and the proposed generalized configuration is trained on Scene~2 only. The role of each curve is detailed in the following.

The genie-aided benchmark is denoted by Actual-CIR-Scene1. Here, we assume that the perfect real-world channel of Scene~1 is available at the BS. The precoder is computed from this perfect channel and is then used to transmit data over the real-world Scene~1 channel for BER evaluation. In practice, however, the BS can never acquire the true channel. Therefore, this curve does not constitute a realizable benchmark but rather a theoretical lower bound on the achievable BER. As expected, it attains the best BER performance over the entire SNR range.

The RT-NN-CIR-Scene1 curve corresponds to the case where the network is trained and deployed in the same environment. We train the U-Net on Scene~1 data by mapping the RT-generated Scene~1 CIR at its input to the real-world Scene~1 CIR at its output. The trained model is then used at the BS to estimate the Scene~1 channel. This estimated channel is used for the precoder design, and data transmission is conducted over the real-world Scene~1 channel to compute the BER. This curve reflects the performance attainable when the model is trained and deployed under matched environmental conditions.

The RT-NN-CIR-Scene2-1 curve demonstrates the generalization capability of the framework across spatially distinct environments. In this case, we train the U-Net only on Scene~2 data by mapping the RT-generated Scene~2 CIR to the real-world Scene~2 CIR. We then deploy the trained model in the unseen Scene~1 environment. During the data transmission phase, the RT engine generates the Scene~1 CIR, which is fed to the trained model to infer a close-to-real-world Scene~1 channel. This inferred channel is used for the precoder design, while the real-world Scene~1 channel is used for the final BER computation. The objective is to demonstrate that a model trained for one scenario can correct the RT-estimated channel and predict the channel for a similar but previously unobserved environment. This capability enables channel prediction and resource allocation without exhaustive site-specific retraining. It is evident from Fig.~\ref{fig5} that the generalized RT-NN-CIR-Scene2-1 configuration incurs a performance gap compared with the matched RT-NN-CIR-Scene1 case. Nevertheless, its BER decreases as the SNR increases and tracks the matched configuration. This observation confirms the spatial transferability of the proposed HF-DT framework.

Fig.~\ref{fig6} shows the SE versus SNR for the same test case. It is evident from Fig.~\ref{fig6} that the genie-aided Actual-CIR-Scene1 benchmark attains the highest SE over the entire SNR range, since it uses the perfect channel for precoding. The matched RT-NN-CIR-Scene1 configuration follows closely. This confirms that the HF-DT estimate supports near-optimal beamforming when the model is trained and deployed in the same environment. The generalized RT-NN-CIR-Scene2-1 configuration incurs a moderate SE gap. Nevertheless, its SE increases steadily with the SNR and tracks the matched configuration. This trend is consistent with the BER behavior in Fig.~\ref{fig5} and further confirms the spatial transferability of the proposed HF-DT framework.

\subsection{Transfer Learning Performance}
In Fig.~\ref{fig7}, we investigate the efficacy of TL in narrowing the performance gap that arises when the model is deployed in a target environment with limited locally available data. The reference curve, RT-NN-Task2, in which the label Task2 refers to test case 2, corresponds to a model jointly trained on the combined datasets of Scene~1 and Scene~2 and tested on Scene~1 without any TL. The RT-NN-TL-C1 curve adopts a TL strategy in which the network is first trained on the Scene~2 dataset and is then fine-tuned through retraining using only $10\%$ of the Scene~1 dataset before being evaluated on Scene~1. The RT-NN-TL-C2 curve follows the same procedure but increases the fraction of Scene~1 data used for fine-tuning to $30\%$. It is evident from Fig.~\ref{fig7} that TL improves the BER performance and enables the fine-tuned models to approach the jointly trained reference. The $30\%$ retraining case attains a closer match, owing to the larger volume of target-domain data incorporated during adaptation. These observations confirm that a model pretrained on a previously mapped environment can be efficiently adapted to a new environment using only a small fraction of locally collected data, which substantially reduces the empirical data collection burden.

Fig.~\ref{fig8} shows the SE versus SNR for the TL test case. The jointly trained RT-NN-Task2 reference attains the highest SE, since it benefits from the full target-domain data. The two TL configurations approach this reference as the SNR increases. It is evident from Fig.~\ref{fig8} that the RT-NN-TL-C2 curve, fine-tuned with $30\%$ of the Scene~1 data, yields a higher SE than the RT-NN-TL-C1 curve, which is fine-tuned with only $10\%$ data. This ordering mirrors the BER results in Fig.~\ref{fig7}. It confirms that a larger fraction of target-domain data during adaptation improves the received SNR and the resulting SE.

\subsection{Neural Receiver Performance}
In Fig.~\ref{fig9}, we quantify the performance gains attainable by integrating the unified neural receiver into the end-to-end communication pipeline. The RT-NN-Task2 curve, reproduced from Fig.~\ref{fig7}, serves as the baseline in which the receiver employs separate demodulation and decoding blocks, with the model jointly trained on Scene~1 and Scene~2 and tested on Scene~1 without TL. The RT-NN-NR curve replaces these disjoint blocks with the proposed neural receiver while retaining an identical system configuration and training procedure. To further isolate the contribution of the digital twin at the transmitter, the Synthetic-NR curve removes the RT and NN blocks at the transmitter and instead generates the channel gain from a Rayleigh fading distribution. This synthetic channel gain is used to compute the precoder, after which data transmission is carried out over the real-world Scene~1 channel for BER evaluation.

It is evident from Fig.~\ref{fig9} that the proposed RT-NN-NR configuration achieves the lowest BER across the entire SNR range. This result confirms that the joint, data-driven receiver compensates for the residual channel impairments and inter-carrier interference more effectively than the conventional cascade of independent demodulation and decoding blocks. Both neural receiver configurations, RT-NN-NR and Synthetic-NR, outperform the RT-NN-Task2 baseline. This observation underscores the substantial detection and coding gains contributed by the unified receiver. Moreover, the pronounced gap between the RT-NN-NR and Synthetic-NR curves highlights the value of the HF-DT generated channel knowledge at the transmitter relative to a statistically assumed Rayleigh channel, since accurate digital twin based channel estimation enables a more reliable precoder design than a generic fading model.

\subsection{Scalability and Interpolation Performance}
To assess the scalability of the proposed framework, we evaluate the temporal interpolation task. In this task, the CIR of the estimation symbols is used to predict the CIR of the remaining symbols within the subframe, following the time slot structure in Fig.~\ref{fig4}.

Table~\ref{tab3} reports the normalized mean squared error (NMSE) of the proposed Transformer-based interpolator against the spline and the LSTM baselines. The NMSE is computed between the predicted CIRs and the true CIRs over the prediction time slots. We report the spline interpolation for two estimation schemes, namely the synthetic Rayleigh-based estimation and the proposed HF-DT estimation. This distinction is made because the spline prediction depends on the underlying estimation scheme. The channels are correlated across time slots. Therefore, the quality of the anchor estimates directly governs the achievable interpolation accuracy. This effect is evident from Table~\ref{tab3}, where the spline scheme attains a substantially lower NMSE under HF-DT estimation ($12.5592$~dB) than under the synthetic estimation ($48.9763$~dB). Among the two baselines that operate on the HF-DT anchors, the LSTM ($11.7770$~dB) and the spline ($12.5592$~dB) attain comparable accuracy. The proposed Transformer is the only scheme that drives the NMSE below $0$~dB, and it attains $-17.2584$~dB. It therefore outperforms the best baseline by approximately $29$~dB. The Transformer and the LSTM contain a comparable number of trainable parameters, and both models are trained on the identical anchor dataset with the same loss function and training protocol. Therefore, the NMSE differences reported in Table~\ref{tab3} reflect the architectural differences between the global attention mechanism and the sequential recurrence, rather than a disparity in the model capacity or the training budget.

\setlength{\textfloatsep}{0pt}
\begin{table}[h]
\vspace{-0.3cm}
\centering
\captionsetup{justification=centering}
\caption{Interpolation NMSE comparison.}
\begin{tabular}{|p{3.0cm}|p{2.1cm}|}
\hline
\bfseries{Interpolation Method} & \bfseries{NMSE [dB]}  \\
\hline
Spline (Synthetic data) & 48.9763  \\
\hline
Spline (HF-DT data) & 12.5592  \\
\hline
LSTM & 11.7770 \\
\hline
\textbf{Transformer} & \textbf{$-17.2584$}  \\
\hline
\end{tabular}
\label{tab3}
\end{table}

\textit{Scalability Results Insights:} The superiority of the Transformer for CIR interpolation arises from its attention mechanism. This mechanism handles the non-linear and highly dynamic nature of radio channels more effectively than sequential or polynomial methods. The spline method assumes local smoothness and relies on static polynomial curve fitting. Hence, it ignores the underlying physics of multipath fading, Doppler shifts, and scattering effects that govern the dynamic CIR. The LSTM model processes the data one time step at a time. This sequential processing introduces a recency bias, which heavily weights the immediately preceding symbols and struggles to retain information from earlier in the sequence. Furthermore, predicting a symbol in the middle of a subframe requires forward-looking context. A bidirectional LSTM can address this limitation. However, it incurs significant latency, since the entire sequence must be processed sequentially in both directions. This latency is a critical drawback under real-time 5G NR constraints. In contrast, the Transformer evaluates the entire time horizon simultaneously. It computes the relevance of every sparse anchor symbol to every missing symbol across the coherence time. The self-attention mechanism allows the model to adjust its interpolation according to the instantaneous channel conditions and the compressed spatial representation learned from the estimation symbols. When the channel exhibits rapid scattering, the model prioritizes the most relevant anchors. When the channel is stable, it averages the anchors more evenly. Because the Transformer identifies the correlation among all time symbols in parallel rather than propagating hidden states across steps, it enables more accurate prediction while still meeting the microsecond latency requirements of the 30~kHz numerology.

\section{Conclusions}

In this paper, we presented an AI-empowered channel twin framework that addresses the computational and latency bottlenecks of integrating high-fidelity 3D digital twins into real-time cellular networks. We proposed the HF-DT framework, in which a U-Net learns the correlation between the RT-generated CIR and the real-world CIR. The simulation results showed that this framework reconstructs the channel accurately from sparse temporal anchors, and the complexity analysis established its feasibility within the 1~ms subframe budget of the 5G NR 30~kHz numerology. We further addressed the challenge of spatial scalability through channel twin generalization. A model trained on one mapped environment was successfully deployed in a nearby unseen environment with similar morphology, without exhaustive site-specific calibration. In addition, we demonstrated that TL narrows the residual performance gap, since fine-tuning with only a small fraction of the target-domain data approaches the jointly trained reference. We also integrated a unified neural receiver that executes demodulation and decoding within a single computational block. This receiver achieved the lowest BER among the considered schemes and reduced the baseband processing overhead while improving resilience against residual channel impairments. The corresponding SE results showed that the improved channel estimates translate into a higher achievable rate over the entire operating range. Finally, the proposed Transformer-based interpolator achieved a markedly lower NMSE than the spline and LSTM baselines, which confirms its scalability for real-time CSI prediction. Overall, the proposed framework bridges the gap between deterministic physics simulations and real-time edge execution, and it provides a scalable foundation for advanced massive MIMO deployments in 6G cellular systems. Extending the validation to a wider set of environment pairs, to the reverse transfer direction, and to morphologically distinct sites remains an interesting direction for future work.

\balance
\bibliographystyle{IEEEtran}
\bibliography{references_updated}
\balance

\end{document}